\documentclass[a4paper]{jpconf}
\pdfoutput=1
\usepackage{graphicx}
\usepackage{amsmath}
\usepackage{color}
\usepackage{subfigure}
\begin{document}
\title{Upgrade of the POLDI diffractometer with a ZnS(Ag)/$^6$LiF scintillation detector read out with WLS fibers coupled to SiPMs}
\author{J.-B. Mosset$^{1}$, A. Stoykov$^{2}$, V. Davydov$^{1}$, M. Hildebrandt$^{2}$, H. Van Swygenhoven$^{1,3}$, W. Wagner$^{1}$}
\address{$^1$Paul Scherrer Institut, Spallation Neutron Source Division, Villigen, Switzerland}
\address{$^2$Paul Scherrer Institut, Laboratory for Particle Physics, Villigen, Switzerland}
\address{$^3$Ecole Polytechnique F\'{e}d\'{e}rale de Lausanne (EPFL), NXMM, Lausanne, Switzerland}
\ead{jean-baptiste.mosset@psi.ch}

\begin{abstract}
A thermal neutron detector based on ZnS(Ag)/${}^{6}$LiF scintillator, wavelength-shifting fibers (WLS) and silicon photomultipliers (SiPM) is under development at the Paul Scherrer Institute (PSI) for upgrading the POLDI instrument, a pulse-overlap diffractometer. The design of the detector is outlined, and the measurements performed on a single channel prototype are presented. An innovative signal processing system based on a photon counting approach is under development. Its principle of operation is described and its performances are evaluated on the basis of a Monte Carlo simulation.
\end{abstract}

\section{Introduction}

\begin{figure}[h]
\begin{minipage}{17pc}
\includegraphics[width=15pc]{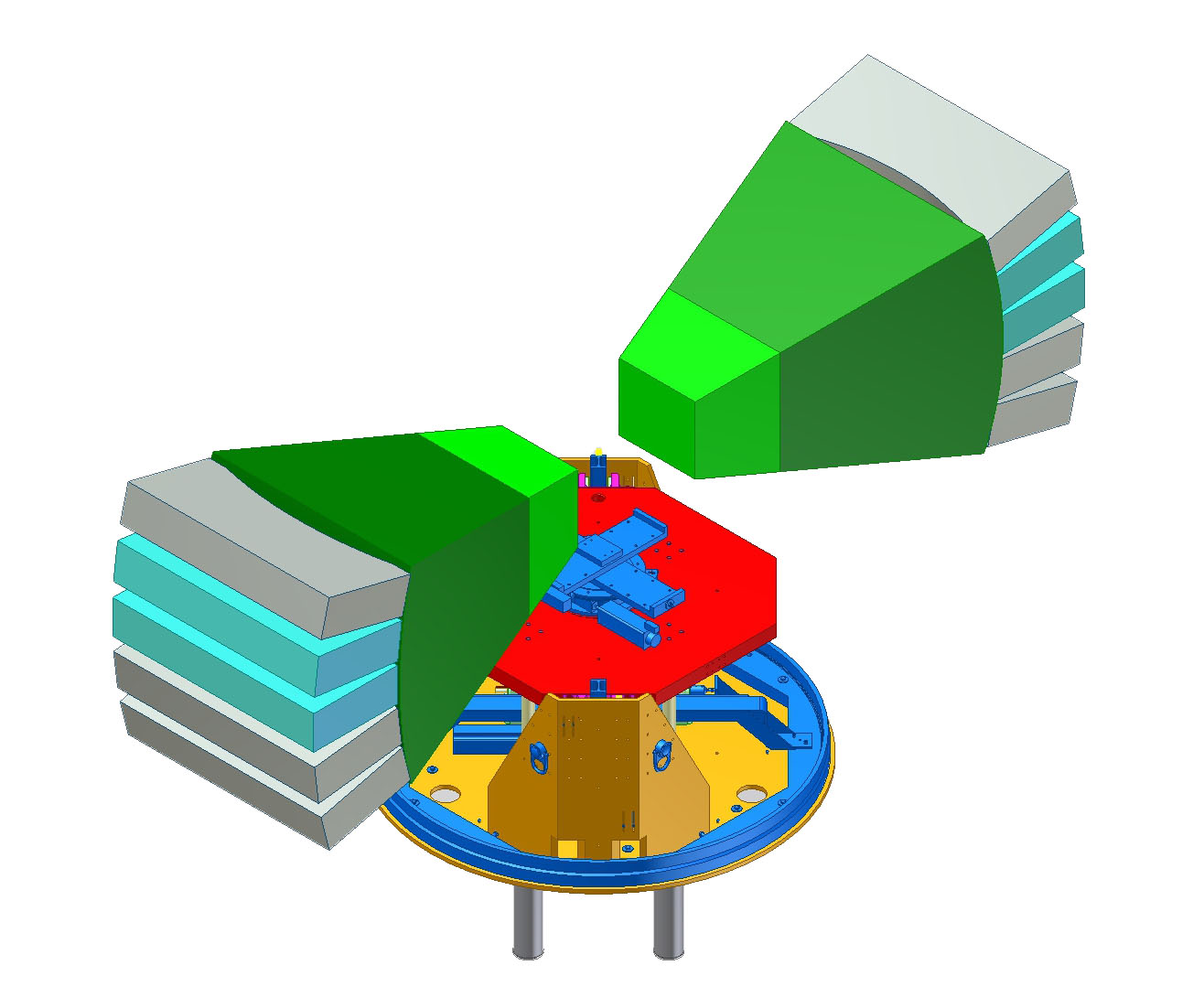}
\caption{View of the aimed new POLDI detector.}
\label{new_poldi_detector}
\end{minipage}\hspace{2pc}
\begin{minipage}{17pc}
\includegraphics[width=15pc]{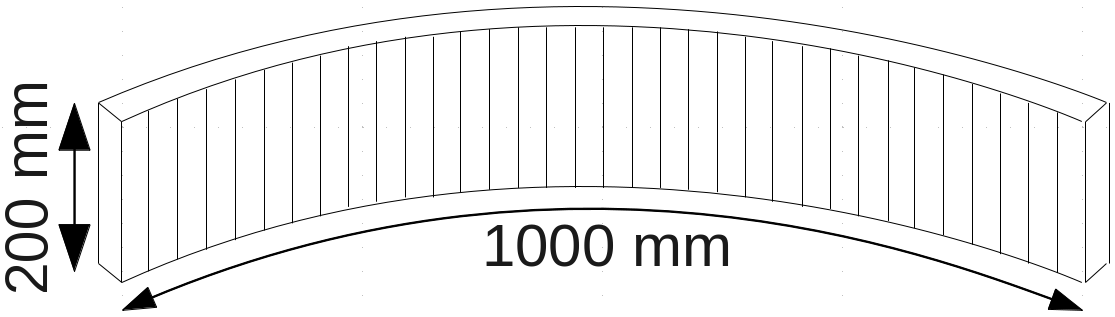}
\caption{Schematic view of the detector module.}
\label{detector_module_geometry}
\end{minipage} 
\end{figure}

POLDI (Pulse-Overlap Diffractometer) \cite{POLDI_2} is a neutron time-of-flight diffractometer commissioned in 2002 at the Swiss neutron spallation source (SINQ) at PSI. The instrument is designed to function as a strain scanner for the investigation of residual stress in engineering materials and components, and for in-situ deformation measurements. To allow a simultaneous measurement of the axial and transverse strain components during in-situ measurements, POLDI will be upgraded with two oppositely placed detector banks based on ZnS(Ag)/$^6$LiF scintillator replacing the actual single $^3$He detector. As shown on Fig.~\ref{new_poldi_detector}, it is foreseen to install five modules per detector bank. However, due to the limited space in the POLDI area, the number of modules could be reduced, depending on the packing fraction of the modules. The minimum requirement is two modules per detector bank.

The actual high price of $^3$He makes the scintillator technology much more attractive in terms of cost. Moreover, from a technical point of view, the scintillator technology is well established and is widely used for diffractometer instruments \cite{status_and_future_neutron_scintillation_detectors,scintillation_detectors}. Unlike the actual detector, the new detector will have a thin detection volume and it will not need to be in time-focusing geometry. This is a clear advantage of the scintillator over the $^3$He technology.

As illustrated in Fig.~\ref{detector_module_geometry}, the detector module has a curved geometry with a 2~m radius of curvature. It is one dimensional position sensitive and its sensitive strips of 200~mm $\times$ 2.5~mm are parallel to its longitudinal axis.  A detector bank consists of 5 detector modules stacked together in a 2~m radius sphere centered on the sample. Table \ref{specifications} shows the specifications of the current and the new detectors.

\begin{figure}[h]
\begin{center}
\resizebox{8cm}{!}{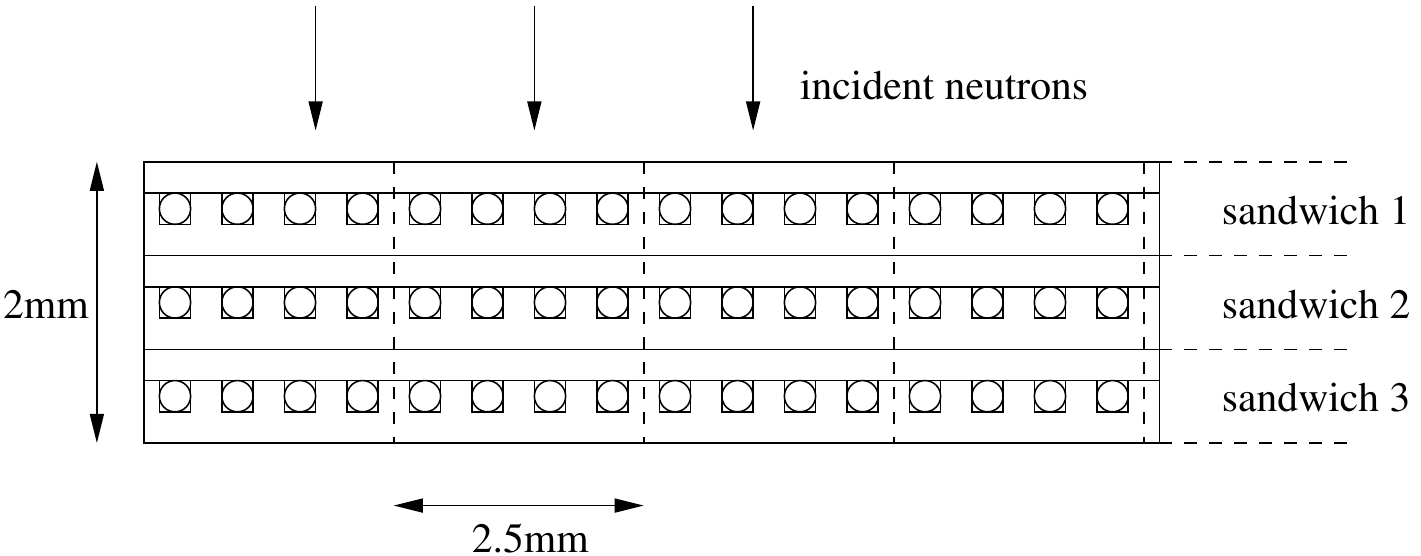}
\caption{Cross-sectional view of a possible detector module. The channels are delimited by the dashed lines. One channel consists of 12 WLS fibers.}
\label{cross_section_detector_module}
\end{center}
\end{figure}

Each single channel of a module may consist of several sandwiches made of an array of wavelength shifting (WLS) fibers squeezed between two ZnS(Ag)/${}^{6}$LiF screens. Fig. \ref{cross_section_detector_module} shows a schematic view of a module made of 3 sandwiches with 12 WLS fibers per channel. The fiber pitch and the thickness of the scintillating layers are geometrical parameters which need to be optimized in terms of light yield and neutron absorption to maximize the neutron detection efficiency. The WLS fibers belonging to the same channel are grouped toghether and coupled to a silicon photomultiplier (SiPM). The use of SiPMs instead of photomultiplier tubes (PMT) is motivated by their high packing fraction and their insensitivity to magnetic fields. The space in the POLDI experimental area is very limited and tests of samples in high magnetic fields are foreseen.

\begin{table}[h]
\begin{center}
\caption{Specifications of the current and new detectors.}
\begin{tabular}{lll}
\br
& actual detector & new detector\\
\mr
Detector technology & ${}^3$He & ZnS(Ag)/${}^{6}$LiF\\
Size of one module & $1 \times 0.2$~m${}^2$ &  $1 \times 0.2$~m${}^2$\\
Number of modules & 1 &  $2 \times 5$\\
Vertical coverage & $\pm 3.2\,^{\circ}$ & $\pm 16\,^{\circ}$\\
Horizontal coverage & $2\theta=75\,^{\circ}$ to $105\,^{\circ}$ & $2\theta=75\,^{\circ}$ to $105\,^{\circ}$\\
Secondary flight path & 2~m & 2~m\\
Angular channels & 400 & 400\\
Spatial resolution & 2.5~mm & 2.5~mm\\
Wavelength & 1.2~\r{A} - 6~\r{A} &  1.2~\r{A} - 6~\r{A}\\
Detection efficiency (at 1.2~\r{A}) & 65\% & 65\%\\
Time resolution & $< 2$~$\mu$s & $< 2$~$\mu$s\\
Sustainable count rate & 4~kHz/ch & 4~kHz/ch\\
Gamma sensitivity & No issue & $<10^{-6}$\\
Intrinsic noise & $< 0.003$~Hz/ch & $< 0.003$~Hz/ch\\
$\Delta$Q/Q resolution & $1\times10^{-3}$ - $2\times10^{-3}$ & $1\times10^{-3}$ - $2\times10^{-3}$\\
\br
\end{tabular}
\label{specifications}
\end{center}
\end{table}

\section{Test of a single channel prototype}

\subsection{Experimental-set-up}

\begin{figure}[h]
\begin{center}
\begin{minipage}{13pc}
\begin{center}
\includegraphics[width=12pc]{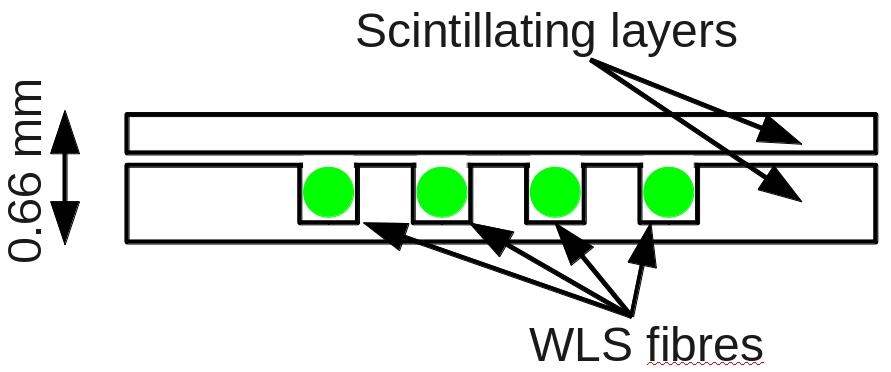}
\caption{Schematic of the sandwich cross-section.}
\label{schematic_prototype}
\end{center}
\end{minipage}
\hspace{0.pc}
\begin{minipage}{11pc}
\begin{center}
\includegraphics[width=10pc]{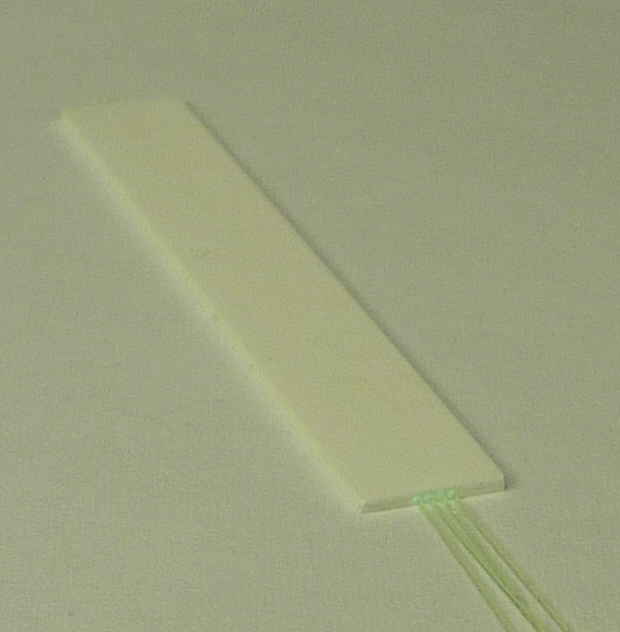}
\caption{The single channel sandwich before polishing of the fiber-end.}
\label{zoom_on_a_sandwich}
\end{center}
\end{minipage}
\hspace{0.pc}
\begin{minipage}{12pc}
\begin{center}
\includegraphics[width=4pc]{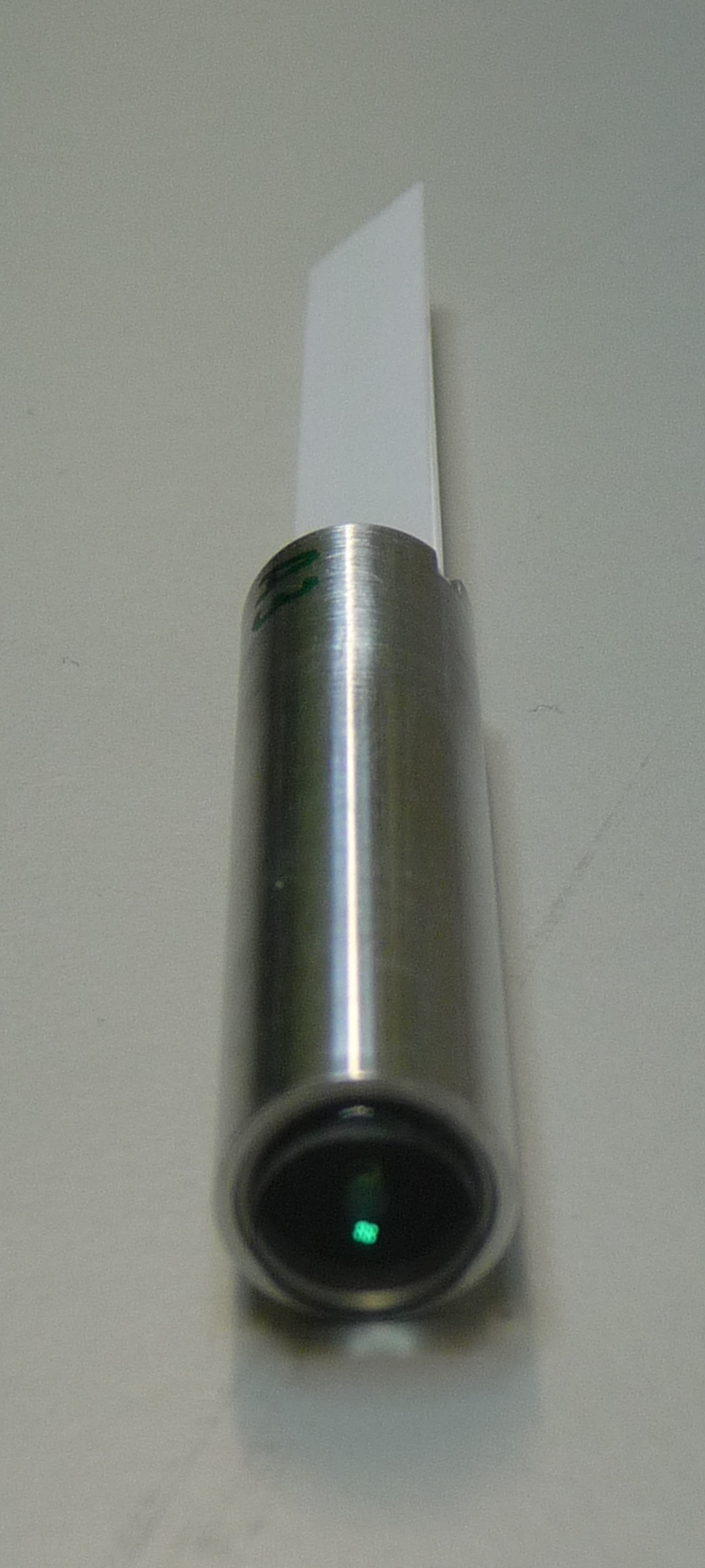}
\caption{The single channel sandwich ready to be connected to a photodetector.}
\label{sandwich_polished}
\end{center}
\end{minipage}
\end{center}
\end{figure}

Fig. \ref{schematic_prototype} shows the single channel prototype which has been tested. It consists of one sandwich made of 2 scintillating layers (ND(2:1) from Applied Scintillation Technologies \cite{applied_scinti_techno}) and four WLS fibers (Y11(200)MC from Kuraray \cite{kuraray}) with a diameter of 250~$\mu$m. The fiber pitch is 0.6~mm. In the 440~$\mu$m thick layer (labelled bottom layer), four grooves of 300~$\mu$m width and 300~$\mu$m depth are machined. The fibers are glued into the grooves with an optical epoxy (EJ500 from Eljen technology \cite{eljen}) and the top scintillating layer 220~$\mu$m thick (labelled top layer) is glued with the same optical epoxy.

On one side, the fibers are cut along the edge of the sandwich and polished. Then, an aluminum foil acting as a mirror is glued to improve the light yield on the other side of the fibers where the photodetector is connected. On this side, the fibers are glued together into the hole of a plexiglas holder and polished. Fig. \ref{zoom_on_a_sandwich} and \ref{sandwich_polished} show a picture of the sandwich before and after polishing of the fiber-end.

\begin{figure}[h]
\begin{center}
\includegraphics[width=25pc]{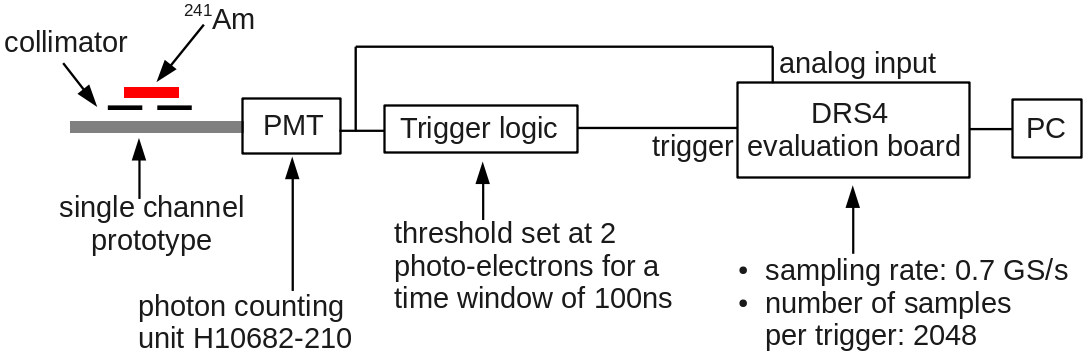}
\caption{Schematic of the experimental setup used to characterize the intrinsic characteristics of a scintillator/fiber sandwich.}
\label{test_setup}
\end{center}
\end{figure}

In order to study the intrinsic characteristics of the scintillator sandwich, the light yield distribution and the decay curve, the fibers were coupled to a low noise photodetector (H10682-210 from Hamamatsu \cite{hamamatsu}), this allowing to set the trigger threshold at a very low level. Fig. \ref{test_setup} shows a schematic of the experimental setup. The sandwich is irradiated with an ${}^{241}$Am source. The energy of $\alpha$ particles emitted by the source (5.48~MeV) is close to the energy left in the scintillator by the triton (2.73~MeV) and $\alpha$ particles (2.05~MeV) resulting from the absorption of a neutron by a ${}^{6}$Li nucleus. The use of an $\alpha$-source is less constraining than the use of a neutron source (higher rate, no need for radiation shielding) and it has the advantage that $\alpha$ particles are stopped close to the scintillator surface, within the first 22~$\mu$m, this allowing to measure the light yield distribution as a function of the distance from the fiber by varying the thickness of the scintillating layer exposed to the source. Furthermore, the higher is the distance between the interaction point and the fibers, the lower is the light yield. So, the light yield measured with $\alpha$ particles is expected to be smaller than the one measured with neutrons, the neutrons being absorbed at any depth in the sandwich. In this sense, the use of an $\alpha$-source provides conservative results.

The partially dead area located on the edge of the sandwich, where there are no fibers, was covered with a 50~$\mu$m thick aluminum foil, to prevent the interaction of $\alpha$ particles in this region. The optical fibers are coupled without glue or grease to the photon counting unit H10682-210 from Hamamatsu. This unit consists of a PMT and a discriminator which triggers at the level of one photoelectron providing a TTL signal on its output. This counting unit has a dark count rate of 50~Hz at 25${}^{\circ}$C, a quantum efficiency of about 20\% at 476~nm (the peak emission wavelength of the fibre) and a pulse-pair resolution of 20~ns.

The output of the photon counting unit is sampled with the DRS4 evaluation board  \cite{DRS4} at a frequency of 0.7 GS/s. For each trigger, 2048 samples are saved, this allowing to study the train of photons over a time window of about 2~$\mu$s. The threshold of the trigger logic is set at 2 photo-electrons for a time window of 100~ns.

\subsection{Results}

\begin{figure}[h]
\begin{center}
\begin{minipage}{17pc}
\begin{center}
\includegraphics[width=15pc]{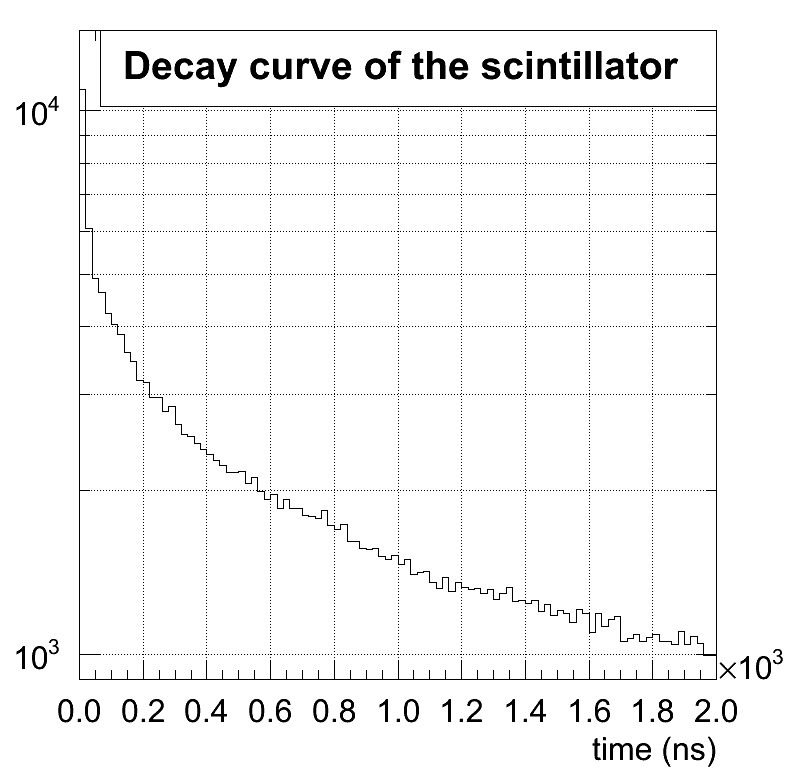}
\caption{Decay curve in a time range of 2~$\mu$s.}
\label{decay_curve_short}
\end{center}
\end{minipage}
\hspace{0.5pc}
\begin{minipage}{17pc}
\begin{center}
\includegraphics[width=15pc]{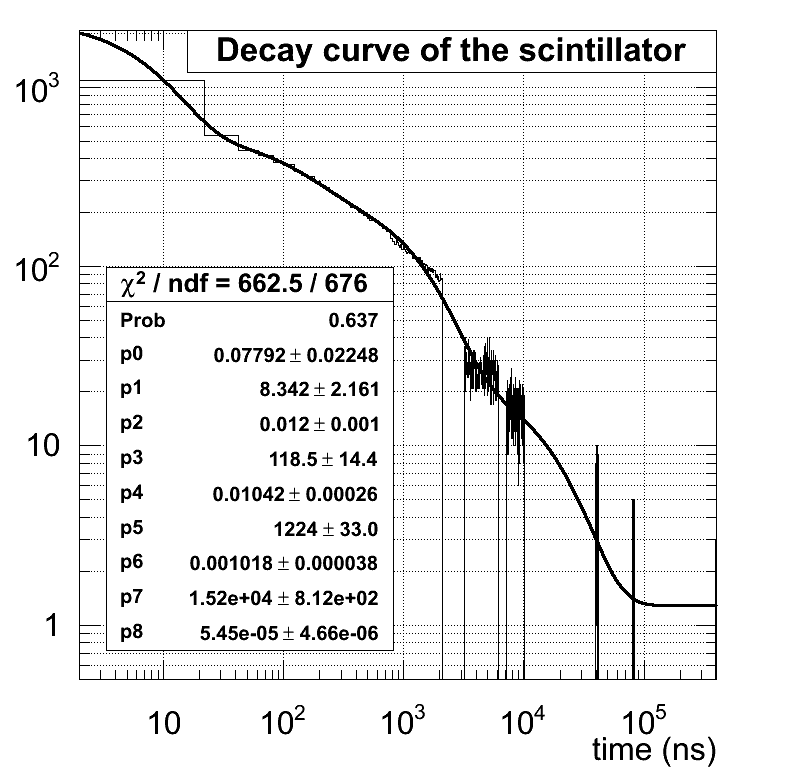}
\caption{Decay curve in a time range of 400~$\mu$s. The solid line is the fitted curve following Eq. \ref{formula_decay_curve}.}
\label{decay_curve_long}
\end{center}
\end{minipage}
\begin{minipage}{17pc}
\begin{center}
\includegraphics[width=15pc]{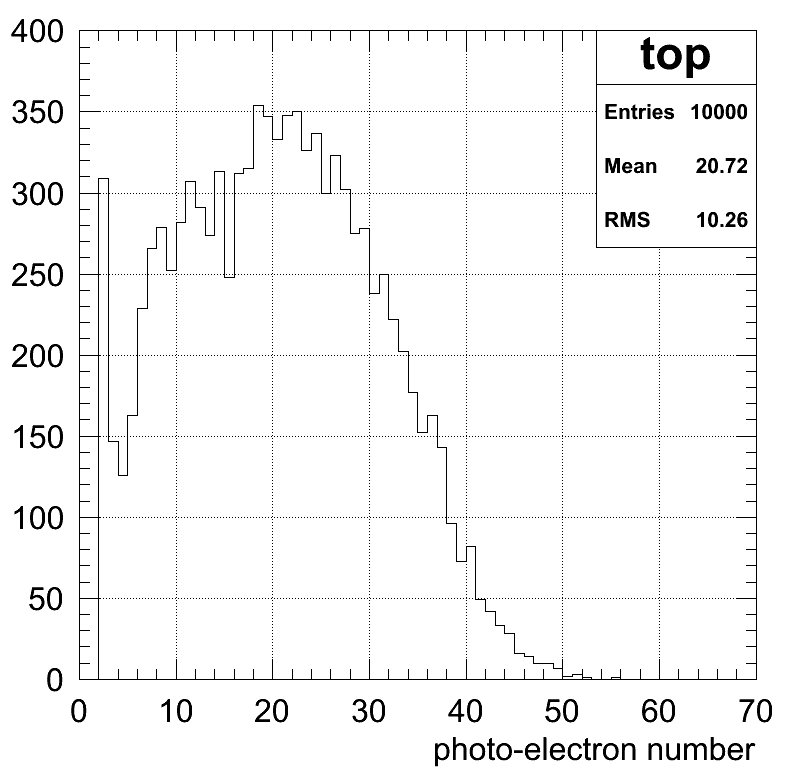}
\caption{Photo-electron spectrum measured when the top layer is irradiated.}
\label{Npe_spectrum_top}
\end{center}
\end{minipage}
\hspace{0.5pc}
\begin{minipage}{17pc}
\begin{center}
\includegraphics[width=15pc]{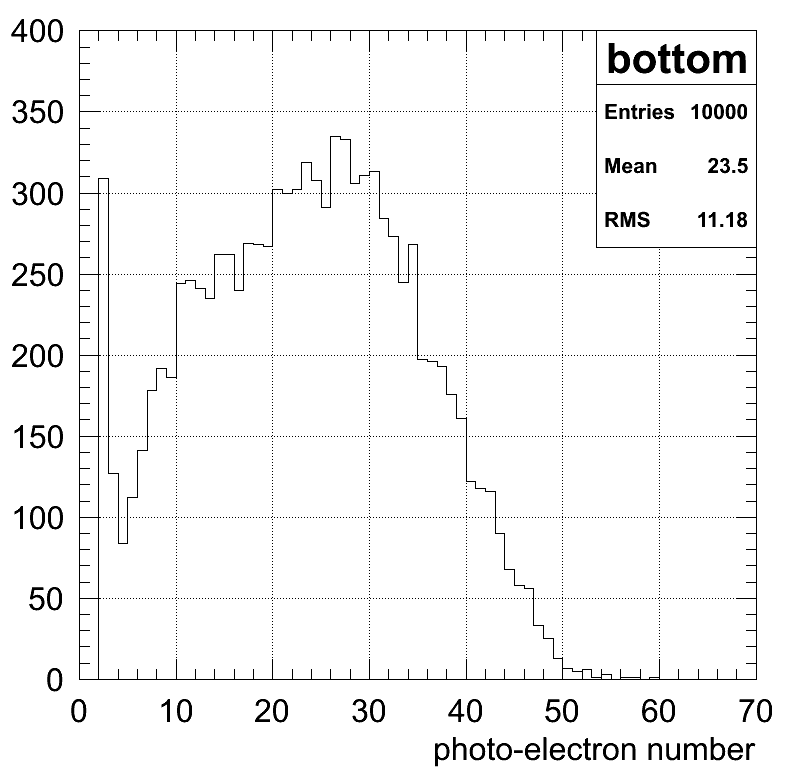}
\caption{Photo-electron spectrum measured when the bottom layer is irradiated.}
\label{Npe_spectrum_bottom}
\end{center}
\end{minipage}
\end{center}
\end{figure}

Fig. \ref{decay_curve_short} shows the decay curve of the scintillator. The limited sampling duration restricts the time range to 2~$\mu$s. To study the slow components of the scintillation, additional measurements are performed with different delays of the trigger: 4, 8, 40, 80 and 400~$\mu$s. Fig. \ref{decay_curve_long} shows the sum of this time spectra (1000 triggers per spectra). In the 0 to 400~$\mu$s time range, the decay curve is analytically well described by the following function:

\begin{equation}
\label{formula_decay_curve}
A(t)=N \cdot (C + \sum_{i=1}^{4} A_i \cdot e^{-t/\tau_i})
\end{equation}

where $N$ is the number of photons measured, $C$ is the background, $A_i$ and $\tau_i$ are the amplitude and the decay time of the component $i$ of the decay. Table \ref{decay_curve_fit} shows the result of the curve fit.

Fig. \ref{Npe_spectrum_top} and Fig. \ref{Npe_spectrum_bottom} show the photo-electron spectrum when the top layer and when the bottom layer are irradiated. The fibers are located 220~$\mu$m away from the surface of the top layer whereas they are only 140~$\mu$m away from the surface of the bottom layer. As expected, the average number of photo-electrons is higher when the bottom layer is irradiated. The peak at 2 photo-electrons appearing on both spectra is due to the afterglow which is still present 20~min after a day light exposure of the scintillator.

\begin{table}[h]
\begin{center}
\caption{Result of the decay curve fitting.}
\begin{tabular}{lllll}
\br
Component & 1 & 2 & 3 & 4\\
\mr
Decay time $\tau_i$ ($\mu$s) & 0.008 & 0.118 & 1.224 & 15.2\\
Amplitude $A_i$ & 0.078 & 0.012 & 0.010 & 0.0010\\
Fraction of photons from the component $i$ ($\frac{A_i \cdot \tau_i}{\sum_{i=1}^{4}A_i \cdot \tau_i}$) & 2\% & 5\% & 41\% & 52\%\\
\br
\end{tabular}
\label{decay_curve_fit}
\end{center}
\end{table}

\section{Development of an innovative signal processing system}

\begin{figure}[h]
\begin{center}
\begin{minipage}{22pc}
\begin{center}
\includegraphics[width=22pc]{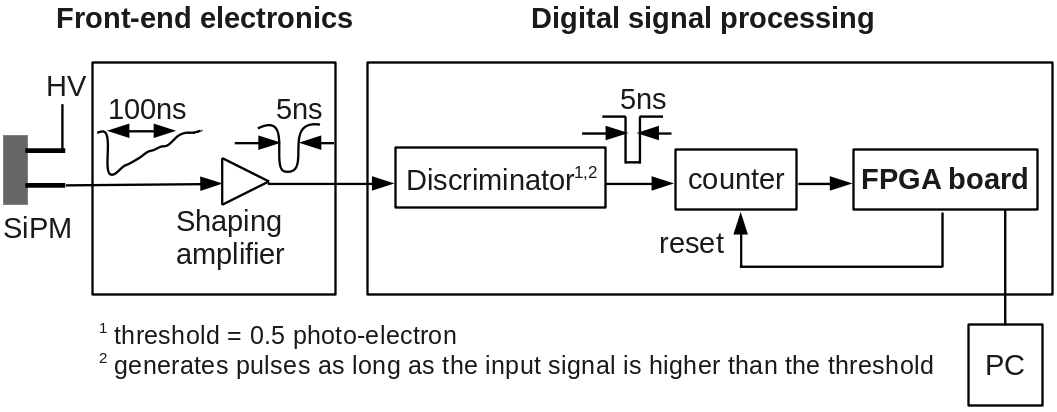}
\caption{Block diagram of the SPS.}
\label{block_diagram1}
\end{center}
\end{minipage}
\hspace{0.5pc}
\begin{minipage}{14pc}
\begin{center}
\includegraphics[width=14pc]{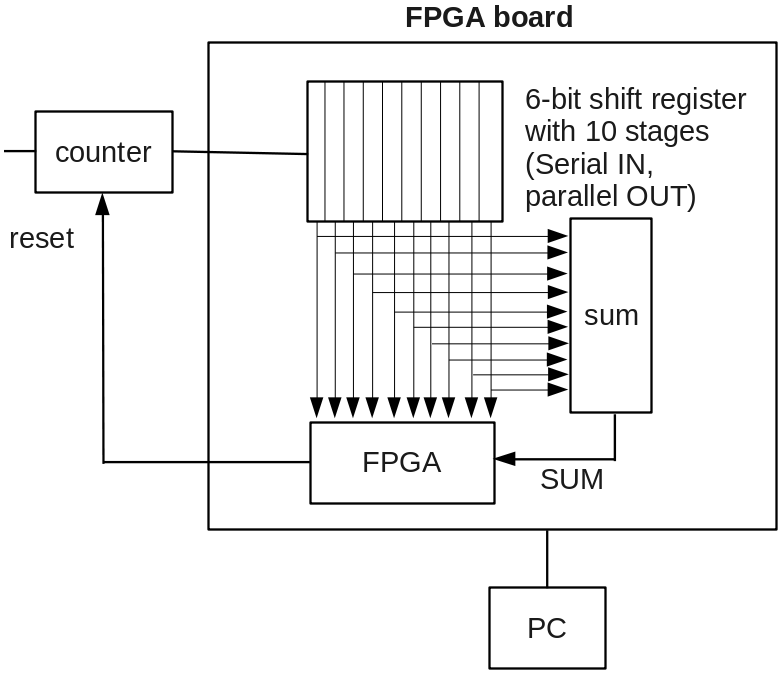}
\caption{Block diagram of the FPGA board.}
\label{block_diagram2}
\end{center}
\end{minipage}
\end{center}
\end{figure}

The signal processing system (SPS) which is under development for the new POLDI detector is based on a photon counting approach. Two reasons makes this approach feasible. Firstly, the ZnS(Ag)/$^6$LiF scintillator is relatively slow and consequently, the photons to detect are sufficiently spaced out in time to allow the detection of almost each individual photon emited in a given time window. Secondly, SiPM have an excellent photon counting capability. Fig. \ref{block_diagram1} and Fig. \ref{block_diagram2} show the block diagrams of the SPS.

The SPS measures (almost) continuously the number of photo-electrons over a time window of 2~$\mu$s. To do that, counting experiments over 200~ns are performed at a frequency of 5~MHz. Every 200~ns, the sum (SUM) of the last 10 values measured by the counter is calculated. If the sum is higher than a threshold value, the readout of the last 10 values is triggered, as well as the next 9 (to be sure to get a sampling of the neutron signal over 2~$\mu$s). Among the 10 samples producing the trigger of the readout, the sample with the maximum count number provides the time stamping of the event. It is called $S_{0}$. The following 9 samples are called $S_{1}$, ..., $S_{9}$. The variable $S_{0}+S_{1}+...+S_{9}$ is used for neutron/noise discrimination and the variable $(S_{0}+S_{1}) / (S_{2}+...+S_{9})$ is used for neutron/$\gamma$ discrimination. Time stamping, n/noise and n/$\gamma$ discriminations are performed online by a FPGA.
\\
\\
The pulse processing algorithm consists of the following sequence of operations:

\begin{enumerate}
\item The FPGA is waiting for $SUM > threshold$ (trigger of the measurement).
\item The FPGA reads 19 samples.
\item The FPGA performs the online analysis: time stamping, n/$\gamma$ and n/noise discrimination.
\item If the event is neutron tagged, the data (time + channel information) are sent to the PC.
\item The FPGA is waiting for $SUM < end_{threshold}$ (end signal). As soon as it measures an end signal, the FPGA return in the state (i).
\end{enumerate}

This signal processing approach offers many advantages compared to a simple signal processing based on a CR-RC pulse shaping network:

\begin{itemize}
\item When several pixels of a SiPM are fired at the same time due to optical crosstalk, only one count is recorded. This allows to get rid of the SiPM crosstalk and this is a crutial point in the reduction of the noise.
\item The pulse shape discrimination becomes straightforward.
\item Numerous parameters of the readout electronics are set in the FPGA firmware, as for example, the clock frequency, the depth of the shift register, the pulse shape criteria. This provides a large flexibility and makes the optimization of the readout electronics easier.
\end{itemize}

\section{Monte Carlo simulation of the signal processing system}

A Monte Carlo simulation has been performed to evaluate the background rejection power and the neutron detection efficiency of the SPS. The input parameters of the simulation are the following:

\begin{itemize}
\item SiPM dark noise: 300~kHz (afterpulses are neglected)
\item neutron rate: 100~Hz
\item decay curve of the scintillator (Fig. \ref{decay_curve_long})
\item photo-electron spectrum (Fig. \ref{Npe_spectrum_top})
\end{itemize}



The Monte Carlo output is a data file which provides, for each 200~ns counting experiment, the number of SiPM dark counts, the number of neutrons absorbed, the number of photo-electrons and the value of $SUM$. From this data file, the detection efficiency and the background rate can be determined as a function of the threshold values for the two variables $SUM$ and $S_{0}+S_{1}+...+S_{9}$.

\begin{figure}[h!]
\begin{center}
\begin{minipage}{17pc}
\begin{center}
\includegraphics[width=17pc]{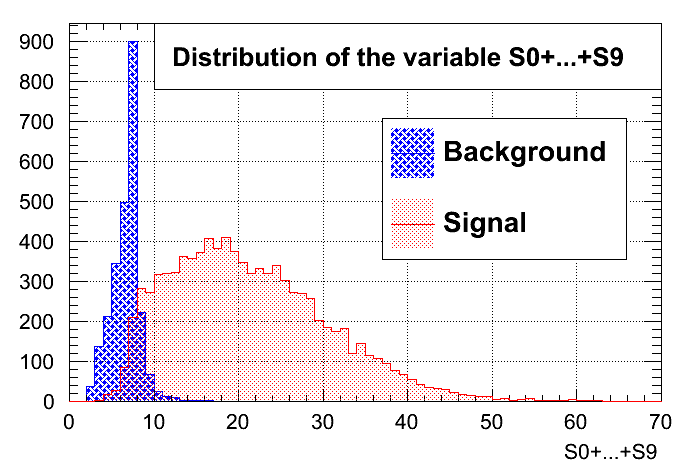}
\caption{Distribution of the variable $S_{0}+S_{1}+...+S_{9}$ for background and for neutron events.}
\label{Stot_spectrum}
\end{center}
\end{minipage}
\hspace{0.5pc}
\begin{minipage}{17pc}
\begin{center}
\includegraphics[width=17pc]{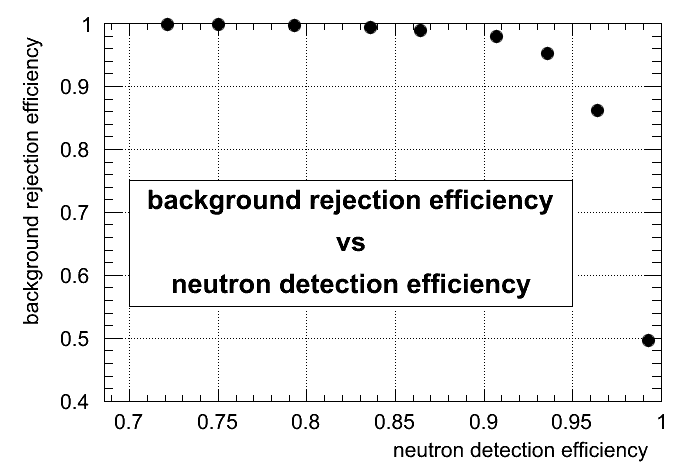}
\caption{Background rejection as a function of the neutron detection efficiency, for events which trigger the readout.}
\label{bg_rej_vs_eff}
\end{center}
\end{minipage}
\end{center}
\end{figure}

For the condition $SUM>6$, the trigger efficiency on neutrons amounts to 90\% and the rate of trigger on the noise amounts to 25~Hz. Fig. \ref{Stot_spectrum} shows the distribution of the variable $S_{0}+S_{1}+...+S_{9}$. The background rejection and the neutron detection efficiency have been calculated as a function of the threshold value for the variable $S_{0}+S_{1}+...+S_{9}$. Fig. \ref{bg_rej_vs_eff} shows the background rejection as a function of the neutron detection efficiency. For a threshold value of 10 on the variable $S_{0}+S_{1}+...+S_{9}$, the background rejection amounts to 98~\% and the neutron detection efficiency amounts to 90\%. It must be stressed that the neutron/noise discrimination is performed only on the events which trigger the readout. Hence, the overall efficiency of the SPS amounts to 81~\% (90\%$\times$90\%) and the background rate amounts to 0.5~Hz (25~Hz$\times$2\%). Preliminary measurements tend to show that the light yield of our test detector is 1.5 times higher when it is exposed to neutrons than when it is exposed to $\alpha$ particles. Hence, the presented results are extremely conservative.


\section{Outlook}
Important next steps are the development of the signal processing system and the test of prototypes read out with SiPMs in a neutron beam. Radiation hardness tests of different SiPMs are underway. The optical isolation between channels or alternatively the processing in the SPS of the optical crosstalk between channels will be investigated.


\section*{References}
\begin{thebibliography}{9}
\bibitem{POLDI_2} U. Stur, ``Time-of-flight diffraction with multiple frame overlap Part II: The strain scanner POLDI at PSI'', {\it Nucl. Instr. and Meth. A}, vol. 545, pp. 330-338 (2005).
\bibitem{status_and_future_neutron_scintillation_detectors} N.J. Rhodes, ``Status and future development of neutron scintillation detectors'', {\it Neutron News}, vol. 17, pp. 16-18 (2006).
\bibitem{scintillation_detectors} N.J. Rhodes, ``Scintillation detectors'', {\it Neutron News}, vol. 23, pp. 26-30 (2012).
\bibitem{applied_scinti_techno} http://www.appscintech.com
\bibitem{kuraray} http://kuraraypsf.jp/index.html
\bibitem{eljen} http://www.eljentechnology.com/
\bibitem{hamamatsu} http://www.hamamatsu.com/jp/en/index.html
\bibitem{DRS4} S. Ritt, ``Design and performance of the 6 GHz waveform digitizing chip DRS4'', {\it Conf. Rec. IEEE Nucl. Sci. Symp.}, pp. 1512-1515 (2008).
\end{thebibliography}
\end{document}